\documentstyle[epsf]{aipproc}
\def\la{\langle} 
\def\ra{\rangle} 
\def\be{\begin{eqnarray}} 
\def\ee{\end{eqnarray}}
\def\zb{\bar{z}}

\newcommand{\eq}{\begin{equation}} 
\newcommand{\eqx}{\end{equation}}
\newcommand{\dl}{\delta}
\newcommand{\eqn}{\begin{eqnarray}} 
\newcommand{\eqnx}{\end{eqnarray}}
\newcommand{\f}[2]{\frac{#1}{#2}}

\newcommand{\Tr}{\mbox{\rm Tr}}

\newcommand{\cor}[1]{\left\langle{#1}\right\rangle}

\renewcommand{\th}{\theta}

\newcommand{\lm}{\lambda}
\newcommand{\chis}{\chi_*}
\newcommand{\chidis}{\chi_5^{dis}}

\newcommand{\nn}{\mbox{\bf{}n}}

\newcommand{\al}{\alpha}

\newcommand{\Pb}{P^\dagger}

\newcommand{\eps}{\varepsilon}
\newcommand{\Gm}{\Gamma}
\newcommand{\om}{\omega}

\begin{document}

\title{U(1)  Problem at Finite Temperature}

\author{ {\bf Romuald A. Janik}$^{1,2}$, {\bf Maciej A.  Nowak}$^{2,3}$,
{\bf G\'{a}bor Papp}$^{4}$ and {\bf Ismail Zahed}$^5$}
\address{$^1$ 
Service de Physique  Th\'{e}orique, CEA Saclay, 91191 Gif-sur-Yvette,
France \\
$^2$ Marian Smoluchowski Institute 
 of Physics, Jagellonian University, 30-059
Krakow, Poland\\ $^3$ GSI, Planckstr. 1, D-64291 Darmstadt, Germany\\ 
$^4$ CNR Department of Physics, KSU, Kent, Ohio 44242, USA \& \\ 
HAS Research Group for
Theoretical Physics, E\"{o}tv\"{o}s University, Budapest, Hungary\\
$^5$Department of Physics and Astronomy, SUNY, Stony Brook, 
New York 11794, USA.}
\maketitle

\begin{abstract}
We model the effects of a large number of zero and near-zero modes in the 
QCD partition function by using sparse chiral matrix models with an emphasis 
on the quenched topological susceptibility in the choice of the measure. At 
finite temperature, the zero modes are not affected by temperature
but are allowed to pair into topologically neutral near-zero modes
which are gapped at high temperature. In equilibrium, chiral and U(1)
symmetry are simultaneously restored for total pairing, evading 
mean-field arguments. We analyze a number of susceptibilities versus 
the light quark masses. At the transition point the topological 
susceptibility vanishes, and the dependence on the vacuum angle 
$\theta$ drops out. Our results are briefly contrasted with recent 
lattice simulations.

\end{abstract}

\section*{Introduction}
The current theoretical resolution of the U(1) problem in QCD relies on 
the assumption that the QCD vacuum supports a finite topological 
susceptibility~\cite{CREWTHER,WITTEN,VENEZIANO}. This assumption
is supported by current lattice simulations \cite{DIGIACOMO}, anomalous 
Ward identities~\cite{WARD}, chiral effective Lagrangians~\cite{EFF} and
canonical quantization~\cite{CANO}, although there are questions in
covariant quantization~\cite{YAZA}. 

In this paper, we will adopt the current view and proceed
to analyze what happens to the U(1) problem at finite temperature. 
At infinite temperature, both the anomaly and topological effects 
become negligible, so that the U(1)
symmetry is effectively restored. In this limit there is an exact 
chiral and U(1) degeneracy modulo quark masses. The question then is 
what happens at finite temperature? Does the U(1) restoration coincide
or differ from the conventional chiral restoration~\cite{SHURYAK1}? 

Recently, a number of lattice simulations~\cite{KOGUT,DETAR,CHANDRA,VRANAS}  
and model calculations \cite{MODELS} have attempted to answer this and other
questions with somewhat opposite conclusions. It is our purpose in this 
letter to try to address some of these issues in a lattice motivated 
matrix model, focusing on the interplay between zero and near-zero modes, 
the effects of light quark masses and the importance of the thermodynamical 
limit. In many ways our analysis will parallel instanton-like calculations
in a solvable context, with interesting lessons for these calculations as well
as lattice simulations. Indeed, one of the main thrust of the present letter
is to provide a minimal framework for a model analysis of current lattice 
results.

In section 2, we motivate the use of a class of chiral 
matrix models by reviewing some recent lattice calculations.
In section 3, we discuss the saddle point results following 
from the present model under some generic assumptions on the
interplay between zero and near-zero modes. In section 4, we 
address certain aspects of the chiral and U(1) transitions,
including a number of susceptibilities. In section 5, we comment
on a number of recent lattice simulations in light of our results.
Our conclusions are in section 6.

\section*{Formulation}
{}
\begin{figure}[htbp]
\centerline{\epsfxsize=110mm \epsfbox{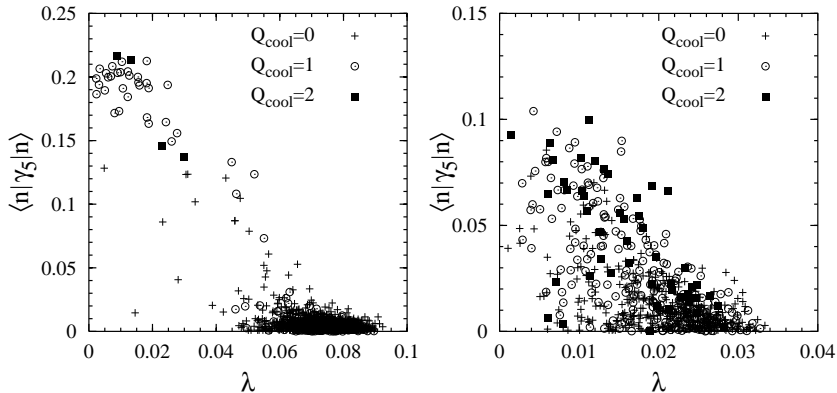}}
\caption{Lattice results from Ref.~\protect\cite{KOGUT} for quenched
$\beta=6.2$ (left) and staggered, $N_f=2$, $\beta=5.55$,
and $m_qa=0.00625$ (right) lattices for topological sectors 0 (crosses),
1 (open circles) and 2 (full boxes).}
\label{lagfig}
\end{figure}
{}
\subsection*{Lattice Motivation}

Recently Kogut, Lagae and Sinclair~\cite{KOGUT} have studied the chirality 
content  $r_n\equiv\cor{n|\gamma_5|n}$ of the low-lying quark eigenstates 
$\lm_n$ for staggered fermions with $D|n\ra=\lm_n|n\ra$. Their results after 
cooling are displayed in Fig.~\ref{lagfig} in the $(r, \lambda )$ plane for 
$\beta =6/g^2=5.55$ and $6.2$, on a $16^3\times 8$ lattice. $|r|=1$ in the 
continuum (about $1/4$ on the lattice) corresponds to an eigenvalue with 
topological charge $\pm 1$, while  $r=0$ corresponds to a non-topological 
eigenvalue. The high temperature configurations ($\beta =5.55$) are 
characterized by a depletion in the zero modes ($r=1/4$), and an enhancement 
in the near-zero modes ($r=0$). Throughout we will refer to the modes with
definite chirality as zero modes, and those without as near-zero modes.
We note that at high temperature and in the continuum, the near-zero modes 
are gapped by $\pm \pi T$.

\subsection*{Model}

A simple way to analyze the interplay between the zero and 
near-zero modes around the chiral transition point is through
a matrix model. A pertinent example for the fermion matrix
$D$, was discussed in~\cite{USNJL} (and references therein)
\eqn
D = 
\left(\begin{array}{cccc}
ime^{i\th}&A\!+\!d&0&\Gm_R^\dagger\\
A^\dagger\!+\!d&ime^{-i\th}&\Gm_L^\dagger&0\\
0&\Gm_L&ime^{i\th}&B\\
\Gm_R&0&B^\dagger&ime^{-i\th}
\end{array}\right) \,,
\label{e.deter}
\eqnx
and the partition function is
\eqn
Z[m,\th ]=
\langle \det D\rangle \,.
\label{e.apart}
\eqnx
For each flavor, 
the entries in the matrices have respectively $n$, $n$, $n_+$ and 
$n_-$ elements corresponding to the number of right handed near-zero modes, 
left handed near-zero modes, right handed zero modes and
left handed zero modes. Here $d$ denotes a diagonal matrix with equal
entries, $d$, the matrices assigned to the near-zero 
modes are square matrices, while the ones assigned to the zero modes
are rectangular matrices. The fluctuations in the rectangularity of the
matrices induce the proper U(1) breaking~\cite{USUA1}. The hopping between 
zero and near-zero modes is characterized by the overlap matrices $\Gamma$.

The averaging in (\ref{e.apart}) is done with respect to the local
fluctuations in the topological charge $\chi =(n_+-n_-)$, with a Gaussian 
width fixed by the quenched topological susceptibility $\chis$,
\eq
\label{e.fluct}
e^{-\f{(n_+-n_-)^2}{2\chis V}} \equiv e^{-\f{\chi^2}{2\chis V}}
\eqx
and a Gaussian measure for the random matrix elements $A, \Gamma, B,$
with width $\Sigma=1$. The latter is physically tied to the quark 
condensate $|\la q^{\dagger} q\ra|\sim (200 \,{\rm MeV})^3$. For simplicity, 
we take the width of the Gaussians to be the same since the
hopping between the low-lying modes may be random enough not to distinguish 
between zero and near-zero modes. The temperature effects on the 
near-zero modes are parameterized by the deterministic and off-diagonal entries 
$d$. They cause the near-zero modes to be gapped by
typically $d=\pm\pi T$ at high temperature, setting the range of
validity of the current assumptions~\cite{USNJL}. The depletion in the
number of zero modes caused by an increase in the temperature will be
discussed below.

We observe that the columns and rows in the fermion matrix (\ref{e.deter})
may be rearranged to give instead
\eq
\left(\begin{array}{cccc}
ime^{i\th}&0&A\!+\!d&\Gm_R^\dagger\\
0&ime^{i\th}&\Gm_L&B\\
A^\dagger\!+\!d&\Gm_L^\dagger&ime^{-i\th}&0\\
\Gm_R&B^\dagger&0&ime^{-i\th}
\end{array}\right)
\eqx
where we grouped the right and left handed modes, respectively.
We denote the total number of modes (size of the full matrix) by 
$2N=2n+n_++n_-$. We note that in the limit $mV\ll 1$ the matrix model 
(\ref{e.apart}) yields sum-rules for the quark eigenvalues that are 
consistent with those discussed in~\cite{SMILGA},
following the general arguments in~\cite{USBIGG} with the identification
$N=\nn V$, where $\nn$ is an order 1 quantity measuring the
density of modes. Physically, the latter sets the scale for the fermionic
contribution to the energy density.

\subsection*{Distribution of Eigenmodes}

The matrices $D$ in (\ref{e.deter}) have (in the chiral limit) a continuous 
distribution of eigenvalues $\tilde{\rho}(\lm)$, with a superimposed
Dirac delta function at zero 
virtuality $\lm=0$, for fixed topological charge $\chi$,
\eq
2N\,\rho(\lm)=|\chi|\dl(\lm)+2N\,\tilde{\rho}(\lm) \,.
\label{dist}
\eqx
To assess the dependence of $r_n$ on $\lm$ one can use the
fact that $\lm_n$ and $\eps r_n$ are the real and imaginary
parts of the non-hermitian operator $D+\eps i{\bf 1}_5$,
in first order perturbation theory. Here ${\bf 1}_5=
{\rm diag} (1, -1,1,-1)$ with each identity assigned to its
pertinent subspace. Taking $\eps$
infinitesimally small and rescaling the imaginary part of the
eigenvalue we obtain the abovementioned dependence\footnote{It is
interesting to note that non-hermitean operators of this
form with $\eps =1/\sqrt{2N}$ yield a generic distribution of non-hermitean
eigenvalues~\protect\cite{EFETOV}.}.

One can now use the chiral structure of $D$ (its anti-commutativity
with ${\bf 1}_5$) to show that the square
\eq
(D+i\eps {\bf 1}_5)^2=D^2-\eps^2
\eqx
is a hermitian operator. This means that the pair $\lm_n+i\eps r_n$ is
either purely real or purely imaginary. It follows that all nonzero 
eigenvalues ($\lm_n^2\gg\eps$) have vanishing $r_n$ while the $|\chi|$ 
topological ones have zero eigenvalue and $r_n=\pm 1$. 
In the limit $mV>1$ the random matrix model (\ref{e.apart})
allows for a model dependent assessment of the distribution of the 
low-lying modes of the Dirac operator in the infinite volume limit
using the methods discussed in~\cite{USCARDANO}.

\section*{Partition Function}

\subsection*{Bosonization}

For equal masses, the partition function~(\ref{e.apart})
can be readily bosonized. The result for fixed size matrices is
\eqn
\label{e.zpart}
Z_{N,n_{\pm}} [m, \theta]=\int dPd\Pb && 
	e^{-NP\Pb} e^{-\f{\chi^2}{2\chis V}} \times\\ 
&&\hspace*{-25mm} 
	\left[(z\!+\!P)(\zb\!+\!\Pb)\!+d^2\right]^n
	(z\!+\!P)^{n_+}\ (\zb\!+\!\Pb)^{n_-}
	\,. \nonumber
\eqnx
Here $z=m e^{i\th}$ stands for degenerate flavors.
The number of near-zero modes $n$, and the number of 
zero modes $n_{\pm}$, fix the size of the
matrices in (\ref{e.zpart}). However, their distribution is
partly fixed by the Gaussian distribution (\ref{e.fluct}), 
while the remaining part is fixed by equilibrium arguments 
as we now discuss.

\subsection*{Detailed balance}

At finite temperature the change in the total number of zero 
modes can be argued generically. Indeed, with increasing temperature the 
zero modes may deplete either by pairing into topologically neutral 
aggregates of near-zero modes~\cite{SHURYAK} or screening~\cite{YAFFE}. 
For the simplest neutral aggregate (molecule~\cite{SHURYAK}), this 
pairing is reminiscent of the Kosterlitz-Thouless one in 
four-dimensions~\cite{ZAHED}.

The chemistry of small neutral aggregates
can be described by the probability of formation $p_f$ and breaking
$p_b$. For molecular arrangements in equilibrium, detailed balance implies

\eq
p_b\, n =p_f\left(\f{n_-}{N}\right)\,n_+ \,.
\label{e.equi}
\eqx
The l.h.s stands for the number of pairs broken.
The r.h.s stands for the number of pairs formed which is the
formation probability, times the 
probability $(n_-/N)$ to find an unpaired negative charge, times
the total number of positive charges $n_+$.  
In equilibrium, the number of pairing matches the number of breaking.

We may now calculate the square of $(n_++n_-)$ from the 
constraint $2N=2n+n_+ +n_-$
and subtract $\chi$ to obtain
\eq
n_+n_-=(N- n)^2-\f{1}{4}\chi^2 \,.
\eqx
Therefore $n$ satisfies
\eq
\f{1}{2} n^2-N(1+\dl)\, n +\f{1}{2}N^2-\f{1}{8}\chi^2=0
\label{addx}
\eqx
with $\dl=p_b/2p_f$. 
Since from~(\ref{e.fluct}) $\chi\sim \sqrt{N}$ we obtain
\eqn
\label{e.npeq}
n = N\underbrace{(1\!+\!\dl\!-\!\sqrt{\dl^2\!+\!2\dl})}_\al -\f{1}{8N}
\f{1}{\sqrt{\dl^2\!+\!2\dl}}\chi^2 + {\cal O}(\chi^4) 
\eqnx
and
\eqn
n_{\pm} = N-n \pm \f{\chi}{2} \,.
\label{e.nmieq}
\eqnx
We are effectively left with a `filling fraction' $\al$ 
and a contribution to the topological susceptibility $\chi$. 
For $\al\to1$ we have $\dl\to0$ ($p_b\ll p_f$). 
In this case, practically all the zero modes are paired (the unpaired
ones are of order $1/\sqrt{N}$)
with U(1) effectively restored~\footnote{Eq.~(\ref{e.npeq}) holds for
$1-\al\gg N^{-1/4}$, hence in this paper the limit $\al\to1$ is
understood always {\em after} the thermodynamical limit $N\to\infty$.}.
For $\alpha\to0$ we have $\delta\to\infty$
($p_b\gg p_f$). In this case, all the zero modes are
unpaired and U(1) is broken.

\subsection*{Saddle point analysis}

We insert~(\ref{e.npeq}-\ref{e.nmieq}) into the
partition function (\ref{e.zpart}), and perform a linear shift in
$\chi$,
\eq
\chi = \tilde{\chi}-i2N\cdot y
\eqx
with the requirement that the term linear in
$\tilde{\chi}$ vanishes in~(\ref{e.zpart}). The resulting consistency
condition (saddle point) reads
\eq
\label{e.consist}
\f12 \log\f{z+P}{\zb+\Pb} +2iay=0
\eqx
where
\eq
a=\f{1}{\chis}\nn
+\f12 \f{\alpha}{1-\alpha^2}\log\f{|z+P|^2+d^2}{|z+P|^2} \,.
\label{defa}
\eqx
The parameter $y$ is just proportional to the average topological charge
\eq
\cor{n_+-n_-}=2V\nn y \,,
\eqx
while $P$ and $\Pb$ in the above equations are the saddle point
solutions following from the 'action'
\eqn
&&%
(1\!-\!\al_*)\log|z\!+\!P|^2\!+\!\al_*\log(|z\!+\!P|^2\!+\!d^2)
-iy\log \f{z\!+\!P}{\zb\!+\!\Pb}
-P\Pb+\f{2\nn}{\chis}\,y^2
\label{effact}
\eqnx
where the effective `filling fraction' is
\eq
\al_*=\al \left(1+\f{y^2}{1-\alpha^2}\right) \,. 
\eqx
Writing out the saddle point equations for $P$ and $\Pb$ and
subtracting yield
\eq
\zb P-z\Pb=2iy \,.
\eqx
This suggests the decomposition $e^{-i\th}P=Q+iy/m$, with $Q$ real,
being the chiral condensate $|\la q^\dagger q\rangle|$ and satisfying the
saddle point equation
\eqn
\label{e.sp}
(1\!-\!mQ\!-\!Q^2\!-\!\f{y^2}{m^2})\,
\big[(m\!+\!Q)^2\!+\!\f{y^2}{m^2}\!+\!d^2\big]=\al_* d^2 \,.
\eqnx
This equation will be analyzed next.

\section*{Results}

The model is totally specified by (\ref{e.deter}-\ref{e.fluct}) 
and (\ref{e.equi}). The thermodynamical limit will be understood
as $N,V\rightarrow\infty$, with $N/V$ fixed.
The parameters are: the width of the Gaussian
$\Sigma=1$, the current mass $m$, the 
quenched topological susceptibility $\chi_*$, 
the deterministic entries $d$, the vacuum angle
$\theta$, the filling fraction $\alpha$ and
the mode-density $\nn=N/V$. Generically, the effects of 
temperature cause $0< d =\pi T$ and 
$0\leq \al \leq 1$.

\subsection*{Chiral condensate}

The $\chi$ saddle point 
equation~(\ref{e.consist}) has a trivial solution, 
$y=0$ for $\th =0$. Inserting this back into the equation
for the condensate $Q$  and setting $m=0$ yields
\eq
Q^2 = \f 12 
\left(1\!-\!d^2 +\sqrt{(1\!-\!d^2)^2+4 (1\!-\!\al)\, d^2}\right)\,.
\label{condx}
\eqx
This result is similar to the one considered by~\cite{SCHAFER},
although our physical interpretation is different. Indeed, in our
case $\alpha$ measures the amount of U(1) breaking and follows from
the rectangular character of the matrices as opposed to the square
matrices used in~\cite{SCHAFER}. It is fixed by detailed balance.
We have ignored the trivial solution with $Q=-m$, by
maximizing the effective action~(\ref{effact}) 

\eqn
F/N = -Q^2 + \log{(m\!+\!Q)^2} + \al \log{%
	\f{(m\!+\!Q)^2+d^2}{(m\!+\!Q)^2}} \,.
\eqnx

For $d<1$, the solution with $Q\neq 0$ sets in independently 
of $\al$. 
For $\al=1$, the zero modes pair into near-zero modes, 
and $Q^2=1-d^2$~\cite{USNJL,SCHAFER,MANYTEMPERATURE}. This is a 
U(1) symmetric phase with broken chiral symmetry. A qualitative 
assessment of the range of temperature where this can take place
follows by reinstating the dimensionful constants,
that is $d=\pi T<\sqrt[3]\Sigma$.
Hence $T<70$ MeV, which is outside the range of validity of our
model (see above). However, this points to the fact that the near-zero
modes are sufficiently gapped at already moderate temperatures,
leaving the zero modes as the only contributors to the chirally
broken phase. Indeed, at $d=1$ we have $Q=\sqrt[4]{1-\alpha}$,
which is zero-mode driven. From here on, only the case with $d>1$ 
will be discussed unless specified otherwise. 

For $\alpha\rightarrow 1$,

\eqn
Q (m) =\f d{\sqrt{d^2-1}}\,\sqrt{1-\alpha} + {\cal O} (m)
\eqnx
while for $\al=1$,

\eqn
Q (m) =     \f m{d^2-1} \left[ 1-\left(\f{d^2}{d^2-1}\right)^3 m^2\right] 
\label{qsolv}
\eqnx
The pairing mechanism suggests an integer `exponent' $\delta=1$.
We recall that for $d=\al=1$, $Q=m^{1/3}$ and $\delta=3$ which 
is mean-field~\cite{USNJL,SCHAFER,MANYTEMPERATURE}.

\subsection*{Isotriplet susceptibilities}

A measure of U(1) breaking in the matrix model can be assessed
by investigating the difference in the $\pi^0$ and $a^0$ isotriplet
susceptibilities~\cite{KOGUT,DETAR}

\eq
\om=\cor{q^{\dagger} i{\bf 1}_5 \tau^3 q q^{\dagger} i{\bf 1}_5 \tau^3 q}_c-
\cor{q^{\dagger} \tau^3 q q^{\dagger} \tau^3 q}_c  
\eqx
and is amenable  to the quark eigenvalue distribution
$\rho(\lm)$ through~\cite{CHANDRA}
\eq
\om=4m^2 \int_0^\infty d\lm \f{\rho(\lm)}{(m^2+\lm^2)^2}\,.
\label{form2}
\eqx
For $\al\rightarrow 1$, the matrix model 
yields $Q\rightarrow 0$ with a gapped spectrum in the
chiral limit, hence $\omega=0$~\footnote{In fact it 
would suffice that $\rho(\lm)$ vanishes as $\sim \lm^a m^b$ 
with $a>-1$ and $a+b>1$.}. This observation is
similar to the one we made in~\cite{USTEMP} without due
care to the U(1) problem as we noted.
In general, $\om$ can be related to the resolvent

\eq
  G(z) = \f 1N \left<\mbox{Tr}\ \f 1{z-D} \right>\,.
\eqx
Specifically,

\eq
\om=\f{{\rm Im\ }G(im)}{m}-{\rm Re\ }G'(im)=\f{Q(m)}{m}-Q'(m)
\label{inter}
\eqx
where $Q(m)$ follows from (\ref{e.sp}).

For $\al < 1$ we have
\eq
\om=\f d{d^2-1}\,\f{\sqrt{1-\al}}{m}+{\cal O}(m^2)
\label{critical}
\eqx
which is to be compared to $\om\sim \sqrt[4]{1-\al}/m$
for $d=1$. The $1/N$ corrections to (\ref{critical}) are 

\eq
2\frac{|\chi|}N\f{1}{m^2}+\f{\chi^2}{N^2}\, (\ldots)\,.
\label{subcritical}
\eqx
The first term is the contribution of the  zero modes
in~(\ref{dist}) through (\ref{form2}).
Since $\chi\sim \sqrt{N}$ both contributions in (\ref{subcritical}) are 
subleading in comparison to (\ref{critical}) in the thermodynamical limit.
These effects may still be present in current lattice assessments of $\om$
as we discuss below.

For $\al =1$, we have

\eq
  \om = \f{d^6}{(d^2-1)^3} m^2 + {\cal O} (m^4) 
\label{critical1}
\eqx
implying that $\omega$ flips from $1/m$ to $m^2$ at the
transition point. We note that $\omega\sim 1/m^{2/3}$
for $d=1$, $\al=1$ which is the mean-field result~\cite{USNJL}. It 
is noteworthy that only integer `exponents' are produced by
the pairing transition, a point in support of some general 
arguments made in~\cite{DETAR}.

\subsection*{Topological susceptibility}

The topological susceptibility in the matrix model
is simply given by

\eq
\chi_{top} = -\frac{\partial^2}{\partial\th^2} \log Z =
-2\f{\partial y}{\partial \th} \,.
\eqx
Expanding the consistency equation to linear order in $y$,
we obtain
\eq
y=\f{-\th}{2a+\f{1}{m(m+Q)}}
\eqx
with $a$ defined in Eq.~(\ref{defa}), which gives
\eqn
\f{1}{\chi_{top}} = \f{1}{\chis} +\sum_{i=1}^{N_f}\f{1}{2m_i(m_i\!+\!Q_i)}
	+\f{\al}{2(1\!-\!\al^2)} \log\f{(m\!+\!Q)^2+d^2}{(m\!+\!Q)^2}
\label{sus}
\eqnx
where we have reinstated the flavor dependence. 
The first contribution is the quenched susceptibility,
the second contribution is the screening caused by the near-zero
modes and the unpaired zero modes, and the third contribution stems 
from the paired zero modes. Note that $\chi_{top}$ vanishes not only
for massless quarks but also for maximal pairing with $\al=1$, as the
asymmetry of $D$'s become minimal. This happens as $Q\rightarrow 0$, 
in qualitative agreement with recent lattice simulations~\cite{DIGIACOMO}.

\subsection*{Pseudoscalar susceptibilities}

The connected and disconnected pseudoscalar susceptibilities
associated with $q^{\dagger} {\bf 1}_5 q$ may be assessed in
a similar way. These susceptibilities were recently addressed
on the lattice~\cite{KOGUT}. In our case, the disconnected 
part $\chidis$ reads

\be
\chidis = 
\frac{1}{N}\cor{\Tr {\bf 1}_5 \f 1{im-D}
{\bf 1}_5 \f 1{im-D}}
\eqnx
and is readily amenable  to (\ref{sus}) through

\eq
\chidis=\f{1}{V}\cor{\f{(n_+-n_-)^2}{m^2}}=\f{\chi_{top}}{m^2}\,.
\label{sus5}
\eqx
In the broken phase $\chi_{top}$ is dominated
by the the second term in (\ref{sus}) for small
$m$, hence $\chidis\sim 1/m$. As $Q\rightarrow 0$,
the limits $m\rightarrow 0$ (chiral) and $\al\rightarrow 1$ 
(pairing) do not commute. For fixed mass 
and $\al\rightarrow 1$, $\chidis\sim (\al\!-\!1)\log{m}/m^2\sim 0$,
while for $\al=1$ and $m\rightarrow 0$, $\chidis\sim 1/(1+Q/m)\sim 1$.
In both cases, $\chidis$ is finite. Note that for $d=1$, $\chidis\sim m^{2/3}$.

The connected part $\chi^{conn}_5$ follows from the 
identity~\cite{USUA1}
\eq
\chi_{top}=\f{2m}{N_f^2}Q-\f{m^2}{N_f^2}\left(
\chi^{disc}_5-\chi^{conn}_5\right) \,.
\eqx
This is the random matrix version of the QCD anomalous Ward
identity~\cite{WARD}. Hence

\eq
\chi^{conn}_5=(N_f^2+1) \chi^{disc}_5-2\f Qm
\eqx
for $m>0$. Again, the connected part of the susceptibility 
is plagued with similar ambiguities in the chiral and pairing 
limits. For $\alpha=1$ and $m\rightarrow 0$, $\chi_5^{conn}$
is finite.

\subsection*{Scalar susceptibilities}

The connected and disconnected isosinglet susceptibility associated
with ${q}^{\dagger}q$ may be estimated in our case as well, following
the lattice conventions~\cite{KOGUT,DETAR}, 

\eqn
\chi^{conn}_S=\frac{1}{N}\cor{\Tr \f{1}{im-D}\f{1}{im-D}}
\eqnx
and
\eqn
\chi^{disc}_S=
\frac{1}{N}\cor{\Tr \f{1}{im-D} \Tr \f{1}{im-D}}
-
\frac{1}{N^2}\cor{\Tr \f{1}{im-D}}^2 \,.
\eqnx     
Both susceptibilities follow from (\ref{e.sp}). Specifically,
\eqn
\chi^{conn}_S&=&Q'(m) = \f 1{d^2-1}\\
\chi^{disc}_S&=&Q^2(m)=\f {m^2}{(d^2-1)^2}\,.
\eqnx
for $\al =1$. This is to be compared with the mean-field result for $d=1$,
$\chi^{conn}_S =1/m^{2/3}$ and $\chi^{disc}_S =m^{2/3}$. 
The factorized result for the disconnected isosinglet susceptibility follows 
from the absence of correlations in the number $(n_++n_-)$.

\subsection*{$\theta$ angle dependence}

In the symmetric phase and for small $m$, the $\theta$ dependence
of the free energy ${\rm ln}Z/V$ is simple. Indeed, since $y\sim m$,
for $\alpha<1$ we may neglect the last term in the consistency equation
(\ref{e.consist}) and obtain
\eq
\sum_i \underbrace{\arctan \f{y/m_i}{Q_i}}_{-\phi_i}=-\th \,.
\eqx
The saddle point equation~(\ref{e.sp}) in the chiral limit
can be solved. Defining $Q_i+iy/m_i\equiv |Q_{i*}| e^{i\phi_i}$,
the result for each flavor is
\eq
\label{e.scale}
\f{y^2}{m_i^2\sin^2\phi_i}= Q^2_{i*} \,.
\eqx
where $Q_{i*}$ follows from $Q$ in (\ref{condx}) through the substitution
$\al\rightarrow \al_*$ for $m=0$. Hence

\eqn
\sum_i \phi_i &=& \th \,,\\
m_1\sin\phi_1=\ldots&=&m_{N_f}\sin\phi_{N_f} \,.
\eqnx
These equations are analogous to the zero-temperature equations
originally derived in QCD~\cite{CREWTHER,WITTEN,VENEZIANO}
and more recently in a matrix model~\cite{USCP}.
Therefore the dependence of the free energy on $\theta$
in the broken phase is the same as the 
vacuum one. The temperature dependence is only
implicit through $Q_*$.


As $Q_*$ approaches zero at the critical point and in the chiral
limit, the dependence on $\theta$ changes. For small $m$, we may
no longer neglect the last term in the consistency equation (\ref{e.consist})
as it diverges. Geometrically the line that intersects the curves of
the $arctan$'s becomes nearly vertical so that $y$ is for all purposes
$0$ regardless of the value of the $\th$ angle. This extends the
result $\chi_{top}=0$ obtained earlier at $\th =0$  to 
$\th\neq 0$.

The fact that the free energy no longer depends on $\th$ at the critical
point and beyond, may be traced to the occurrence of a non-analytic term 
$|\chi|$ in the partition function. Indeed from (\ref{addx}) and 
for $\al=1$
\eq
n =N-\f{1}{2}|\chi| \,.
\eqx
Inserting this into the partition function (\ref{e.zpart}), we obtain
\eq
e^{i\th\chi}e^{-b|\chi|}e^{-\f{\chi^2}{2\chis V}}
\eqx
with $b$ a {\em positive} factor stemming from (\ref{e.zpart}).
Performing the integral/sum over $\chi$ gives
a vanishing contribution to the free energy $\log Z/V$. Specifically, 
the sum over $\chi$ is for a range of parameters well approximated by
\eq
2 {\rm Re\ }\f{1}{1-e^{i\th-b}}
\eqx
which gives zero contribution to the free energy. 
This is a direct consequence of the total quenching of the
topological fluctuations in the paired configurations of zero modes.
The simultaneous restoration of chiral and U(1)
symmetry at finite temperature yields a symmetric phase that preserves 
strong CP.

\section*{Comparison to Lattice}

In a first lattice study by Bernard et al.~\cite{DETAR}, chiral
symmetry restoration was found to precede the U(1) restoration. Their
analysis relied on gauge configurations at fixed lattice spacing
$a\sim 1/6T_c\sim 0.25$ fm~\cite{DETAR} for $N_t=6$. Since finite
volume effects were not investigated, it may be that the small U(1)
breaking effects detected in these simulations through a lattice measurement
of $\omega$ for staggered fermions are of the type (\ref{subcritical}). 
However, simple estimates based on their numbers appear to be on the larger 
side of their reported results~\cite{USCARDANO}. As we already noted,
the pairing mechanism supports integer `exponents' for $\om$, a point
sought in~\cite{DETAR}.

In a second lattice study by Kogut et al.~\cite{KOGUT}, the low-lying 
quark eigenvalues of the staggered Dirac operator where investigated.
Their analysis shows that the disconnected isosinglet susceptibility 
$\chidis$, decreases but remains finite in the high temperature phase.
The finite result was shown to follow from the eigenmodes with finite
chirality (topological). The conclusion was that the U(1) symmetry was
not restored in the symmetric phase, although again finite volume effects
were not investigated. In the present matrix model, we have observed that
$\chidis$ remains finite in the chiral and U(1) symmetric phase for $d>1$,
when the thermodynamical limit is carried. Also, we have noted an ambiguity 
in the limits $m,\al\rightarrow 0,1$, suggesting that the cooling procedure
may be subtle while carrying the chiral limit. Indeed, lattice cooling
affects the ``filling fraction'' $\al$.

In a third lattice study by Chandrasekharan et al.~\cite{CHANDRA},
the chiral condensate and $\omega$ were calculated using also staggered
fermions for fixed $\beta=5.3$ and $N_t=4$. Although their results were
found to be consistent with those of Bernard et. al.~\cite{DETAR}, they
concluded that the anomalous effects were small, hinting at the possible
restoration of U(1) in the symmetric phase. Although their conclusions
are closer to ours in spirit, they differ in content
since their small value of $\omega$ was obtained from a linear extrapolation 
in the current quark mass, as opposed to a quadratic extrapolation suggested
by our results. Also, we have observed that the $\th$-dependence drops in the 
symmetric phase in distinction to a general assumption they made.

In a fourth lattice study by Vranas et al.~\cite{VRANAS},
lattice simulations with domain wall fermions were carried at $N_t=4$.
It was found that the high temperature phase preserves chiral symmetry with 
a small amount of U(1) breaking, although with a somehow heavier pion mass.
The method preserves flavor symmetry and incorporates the effect
of the anomaly at every stage of the simulation.
It is indeed encouraging that the results of these 
simulations are the closest to ours.

\section*{Conclusions}

We have used a simple matrix model to analyze the interplay 
between zero and near-zero modes at finite temperature. While the
model finds its motivation in the lattice results described above,
it was originally argued from an NJL model with U(1) breaking~\cite{USNJL}. 
At finite temperature, the pairing mechanism at work in the
zero mode sector is reminiscent of the one originally
suggested in the context of instantons~\cite{SHURYAK}. The present
model is by no means exhaustive as additional effects, e.g. Debye 
screening, have been omitted. Their consideration goes beyond the scope of 
this work.

This notwithstanding, our results indicate that chiral and U(1) 
symmetry are simultaneously restored for maximum pairing of zero modes. 
Although the chiral condensate receives contribution from all low-lying modes, 
its depletion to zero requires that the zero modes are paired and the 
near-zero modes are gapped. A simple estimate shows that the near zero modes
are substantially gapped at moderately low temperatures, suggesting their
early decoupling. This rules out the possibility of a U(1) restoration prior 
to a chiral restoration, and suggests that both symmetry restorations
occur simultaneously.

The transition by pairing the topological charges is followed
by a number of observations regarding the topological, scalar 
and pseudoscalar susceptibilities for small current quark masses.
In particular, integer `exponents' were observed in contrast to the
fractional exponents expected from general universality arguments.
These susceptibilities have been extensively studied on the lattice.
Our comparison with the most recent lattice simulation using domain
wall fermions is very encouraging, although some improvements regarding 
the extrapolation to zero quark mass and finite volume effects are still 
warranted in the staggered simulations. In many ways, our results should 
benefit the more complex instanton calculations when they become available.

Finally, we have shown that in the symmetric phase the topological 
susceptibility  vanishes in the thermodynamical limit.
As a result, the partition function develops a non-analyticity in 
the net topological charge that causes the symmetric phase to be
CP even whatever the vacuum angle. While admittedly this is a result of
the present matrix model, it should be interesting to see whether it
carries to QCD in the infinite volume limit.

\section*{Acknowledgments}

We would like to thank S. Chandrasekharan for a discussion.
IZ thanks Norman Christ for several discussions, and C.-R. Ji and D.-P. Ming
for the invitation to a pleasant meeting. 
This work was supported in part by the US DOE grants DE-FG-88ER40388
and DE-FG02-86ER40251, by the Polish Government Project (KBN) grant
2P03B00814 and by the Hungarian grant OTKA-T022931.

\end{document}